# The probabilities of transitions during stimulated recombination of atoms

M.A. Kutlan

Institute for Particle & Nuclear Physics, Budapest, Hungary

kutlanma@gmail.com

The probabilities of transitions of the system to the different final states are determined by the values of the amplitudes of the corresponding individual states during stimulated recombination of atoms.

## 1. INTRODUCTION

Laser spectroscopy of hyperfine-structure (HFS) states is being applied in investigations of the properties of atoms and nuclei [1,2 and references therein]. The polarization of nuclei that arises in multiphoton resonance ionization of atoms was considered. In a similar formulation, the problem of the polarization of the photoelectrons that appear as a result of resonance two-photon ionization of unpolarized alkali-metal atoms was solved too.

In connection with experiments on electron cooling of ions in storage rings [3] it is of interest to consider the inverse process - namely, stimulated recombination of atoms (ions) with subsequent population of HFS components. Effects involving stimulated recombination of atoms in proton - electron beams were considered in [4] and [5]. It was shown that under certain conditions, ensured by the method of electron cooling, the rate of stimulated recombination considerably exceeds the rate of spontaneous recombination. In view of this, it is natural to expect that in the case of stimulated recombination in the field of a circularly polarized wave it will be possible to observe effects involving the optical polarization of nuclei.

## 2. THE BASIC EQUATIONS

In the center-of-mass frame of the recombining particles we shall describe their interaction with the first wave by the operator (the wave Propagates along the z axis and is circularly polarized in the xy plane; $\hbar = c = 1$ )

$$V^{(1)}(t) = \frac{eE_{01}}{2}\left(\frac{8\pi}{3}\right)^{1/2} rY_{1\pm1}(\theta,\varphi)\exp(i\omega_1 t) + c.c. \tag{1}$$

where $E_{01}$ is the amplitude of the electric-field intensity and $\omega_1$ is the frequency of the wave (for definiteness, we later confine ourselves to the case of a wave with right polarization; see Fig. 1 of [2]).

With regard to the second wave, which leads to induced 3p-2s transitions, we shall assume that it is quasimonochromatic, with frequency width $\Delta\omega_2$ satisfying the condition $\Delta\omega_2 \geq \Delta\varepsilon_{HFS}$. This inequality originates from the fact that, for typical beam parameters in the method of electron cooling, the Doppler width $\Gamma_{D\|}$ of the 3p level for emissions in the direction along the beams attains the magnitude of the energy interval $\Delta\varepsilon_{HFS}$ between the components F = 1 and F= 0 of the HFS of the 2s level. Thus, the condition $\Delta\omega_2 \geq \Delta\varepsilon_{HFS} \sim \Gamma_{D\|}$ makes it possible to use the recombined atoms more effectively during the subsequent polarization of the nuclei.

At first sight, it might appear admissible to have a degree of nonmonochromaticity of the wave such that $\Delta\omega_2 \geq \Delta\varepsilon_{FS}$ where $\Delta\varepsilon_{FS}$ is the energy of the fine splitting of the components $3p_{3/2}$ and $3p_{1/2}$. In this case, both components of the fine structure of the 3p level would participate, on an equal basis, in the stimulated population of the HFS components of the 2s level. However, as will be shown below, the resonance condition for recombination of atoms leads to the result that this process effectively proceeds only through one component of the fine structure of the intermediate level (under the condition that for the first wave the spectral width $\Delta\omega_1 < \Delta\varepsilon_{FS}$).

Certain restrictions, associated with the condition for optimization of the effects to be observed, are also imposed on the intensities of the waves used. It is necessary that the ionization width $\Gamma_i$ and field width $\Gamma_i$ of the 3p level in the resonance waves be of the order of the total width $\Gamma_n = \gamma_n + \Gamma_{D\|}$ of the level.

We shall describe the interaction of an atom with the second wave, which we shall assume to be circularly polarized, by the operator

$$V^{(2)}(t) = 2\pi^2 e \left(\frac{2}{V\omega_2^2} I_{n,\omega_2}\right)^{1/2} \left(\frac{8\pi}{3}\right)^{1/2} r Y_{1\mp 1}(\theta,\varphi)\exp(i\omega_2 t) + c.c. \qquad (2)$$

where $I_{n,\omega_2}$ is the angular spectral density of the radiation intensity in the wave, normalized by the condition

$$\iint I_{n,\omega_2} d\omega_2 d\Omega_n = I_0$$

($I_0$ is the total intensity of the second wave); V is the normalization volume. We draw attention to the fact that, since we henceforth confine ourselves to an *s* state of the incident electrons in the continuous spectrum, and the final state of the recombined atom is $2s_{1/2}$, the field-intensity vectors in the first and second waves rotate in opposite directions. We note also that the second wave can be linearly polarized along any direction lying in the plane perpendicular to the z axis.

To solve the problem of the behavior of the system in the fields of two resonance waves it is appropriate to use the method of Heitler [6], which leads to the following system of equations for the Fourier transforms of the amplitudes of the individual states:

$$(E-E_0)C_{\mathbf{p}}(E) = 1 + \sum_{n_1,k} V^{(1)}_{\mathbf{p}/n_1} \tilde{C}_{n_1}(E) C_{\mathbf{p}}(E) \zeta(E-E_1) + \sum_{n_2,k} V^{(1)}_{\mathbf{p}/n_2} \tilde{C}_{n_2}(E) C_{\mathbf{p}}(E) \zeta(E-E_2),$$

$$(E-E_1+1/2i\Gamma_{n_1}+1/2i\Gamma_{in_1})\tilde{C}_{n_1}(E)\zeta(E-E_1) = V^{(1)*}_{n_1/p} + \sum_{m_1,k_1} V^{(2)}_{n_1/m_1 k_1} \tilde{C}_{n_1}(E)\zeta(E-E_3)$$

$$+\sum_{m_2,k_1'} V^{(2)}_{n_1/m_2 k_1'} \tilde{C}_{m_2 k_1}(E)\zeta(E-E_4),$$

$$(E-E_2+1/2i\Gamma_{n_2}+1/2i\Gamma_{in_2})\tilde{C}_{n_2}(E)\zeta(E-E_2) = V^{(1)*}_{n_2/p} + \sum_{m_1,k_2} V^{(2)}_{n_2/m_1 k_2} \tilde{C}_{m_1 k_2}(E)\zeta(E-E_5)$$

$$+\sum_{m_2,k_1'} V^{(2)}_{n_2/m_2 k_2'} \tilde{C}_{m_2 k_2'}(E)\zeta(E-E_6), \tag{3}$$

$$\tilde{C}_{m_1 k_1}(E) = V^{(2)*}_{m_1 k_1/n_1} \tilde{C}_{n_1}(E)\zeta(E-E_1),$$

$$\tilde{C}_{m_2 k_1'}(E) = V^{(2)*}_{m_2 k_1'/n_1} \tilde{C}_{n_1}(E)\zeta(E-E_1),$$

$$\tilde{C}_{m_1 k_2}(E) = V^{(2)*}_{m_1 k_2/n_2} \tilde{C}_{n_1}(E)\zeta(E-E_2),$$

$$\tilde{C}_{m_2 k_2'}(E) = V^{(2)*}_{m_2 k_2'/n_2} \tilde{C}_{n_1}(E)\zeta(E-E_2).$$

In Eqs. (3) we have used the following notation: $C_{\mathbf{p}}(t)$ is the amplitude of the system in the continuous spectrum with energy $E_0 = \varepsilon_{\mathbf{p}}$ ( the initial state, with incident-electron energy $\varepsilon_{\mathbf{p}}$ and initial condition $C_{\mathbf{p}}(0)=1$); $\tilde{C}_{n_1}$ and $\tilde{C}_{n_2}$ are the amplitudes of the intermediate bound states of the system (with energies $E_1 = E_{n_1} + \omega_k$ and $E_2 = E_{n_2} + \omega_{k'}$ respectively) that are formed from the initial state under the action of the perturbation $V^{(1)}$ (see Fig. 1 of [2]) ; $\tilde{C}_{m_1 k_1}$ and $\tilde{C}_{m_2 k_1'}$ are the amplitudes of the final states (with energies $E_3 = E_{m_1} + \omega_k + \omega_{k_1}$ and $E_4 = E_{m_2} + \omega_k + \omega_{k_1'}$ respectively) corresponding to the HFS- components of the $2s_{1/2}$ level (the amplitudes $\tilde{C}_{m_1 k_2}$ and $\tilde{C}_{m_2 k_2'}$ and energies $E_5$ and $E_6$ have an analogous meaning); $V^{(2)}_{n_1/m_1 k_1}$ and $V^{(2)}_{n_1/m_2 k_1'}$ are the matrix elements for the stimulated recombination from the s state of the continuous spectrum to the states $3p_{3/2}$ and $2p_{1/2}$, respectively; $V^{(2)}_{n_1/m_1 k_1}$ and $V^{(2)}_{n_1/m_2 k_1'}$ are the matrix elements for the stimulated transitions of the atom in the field of the second wave to the HFS components of the $2s_{1/2}$ level; $\Gamma_{n_1}$ and $\Gamma_{n_2}$ are the total widths of the $3p_{3/2}$ and $3p_{1/2}$ levels, including the natural and Doppler widths; $\Gamma_{in_1}$ and $\Gamma_{in_2}$ are the photoionization widths of the corresponding levels in the field of the first wave; $\zeta(x) = P/x - i\delta(x)$. The system of equations (3) was obtained in the resonance approximation, which assumes fulfillment of the standard conditions: The total widths of the states taking part in the transition, and also the frequency detunings of the waves from resonance, are smaller than the energy distances to the nearest levels and wave frequencies.

The probabilities of transitions of the system to the different final states are determined by the values of the amplitudes of the corresponding individual states." Solution of the system of equations (3) leads to the following expressions for the probabilities of a transition in unit time to states in which the nuclear spin has a specified projection M along the quantization axis (the z axis) [7]:

$$w\left(M = \frac{1}{2}\right)$$
$$= (2\pi)^2 \frac{1}{3}\alpha|\langle 2s|r|3p\rangle|^2 \left\{ \frac{|V^{(1)}_{p/n_1}|^2}{(\varepsilon-\varepsilon_{01})^2 + \tilde{\Gamma}^2_{n_1}/4} \times \left[\left(-2^{1/2} + \frac{1}{3}\right)^2 I_{\tilde{\omega}_{20}} + \frac{1}{9} I_{\tilde{\omega}_{20}'}\right] \right.$$
$$\left. + \frac{4}{9}\frac{|V^{(1)}_{p/n_2}|^2}{(\varepsilon-\varepsilon_{02})^2 + \tilde{\Gamma}^2_{n_2}/4}\left(I_{\tilde{\omega}_{20}} + I_{\tilde{\omega}_{20}'}\right)\right\}, \qquad (4)$$

$$w\left(M = -\frac{1}{2}\right) =$$
$$= (2\pi)^2 \frac{1}{3}\alpha|\langle 2s|r|3p\rangle|^2 \left\{ \frac{|V^{(1)}_{p/n_1}|^2}{(\varepsilon-\varepsilon_{01})^2 + \tilde{\Gamma}^2_{n_1}/4} \times \left[\left(-1 + \frac{2^{1/2}}{3}\right)^2 I_{\tilde{\omega}_{20}} + I_{\tilde{\omega}_{20}'}\right] \right.$$
$$\left. + \frac{8}{9}\frac{|V^{(1)}_{p/n_2}|^2}{(\varepsilon-\varepsilon_{02})^2 + \tilde{\Gamma}^2_{n_2}/4} I_{\tilde{\omega}_{20}}\right\}, \qquad (5)$$

Where $\varepsilon \equiv \varepsilon_p$; $\varepsilon_{01} = \omega_1 - |E_{n_1}|$ and $\varepsilon_{02} = \omega_1 - |E_{n_2}|$ are parameters specifying the amount by which the energy $\omega_1$ of a quantum of the first wave exceeds the threshold for ionization from the

corresponding level; $\langle 2s|r|3p\rangle$ is the radial matrix element of the $3p \to 2s$ transition; $I_{\omega_{20}}$ and $I_{\tilde{\omega}_{20}}$ are the spectral densities of the second wave, taken at the frequencies $\omega_{20} = \varepsilon - E_{m_1} - \omega_1 \approx \Delta E_{n_1 m_1}$ [the transition $3p_{3/2} \to 2s_{1/2}$ (F = 1)] and $\tilde{\omega}_{20} = \varepsilon - E_{m_2} - \omega_1 \approx \Delta E_{n_2 m_2}$ [the transition $3p_{1/2} \to 2s_{1/2}$ (F= 1)]; $\omega'_{20} \approx \omega_{20} + \Delta\varepsilon_{HFS}$ and $\tilde{\omega}'_{20} \approx \tilde{\omega}_{20} + \Delta\varepsilon_{HFS}$; $\tilde{\Gamma}_{n_1}$ and $\tilde{\Gamma}_{n_2}$ are the total widths of the $3p_{3/2}$ and $3p_{1/2}$ levels, including the natural width $\gamma_n$, the Doppler width $\Gamma_{D\|}$ the photoionization width $\Gamma_i$ in the field of the first- wave, and the field width $\Gamma_f$ in the field of the second wave: $\tilde{\Gamma}_n = \gamma_n + \Gamma_{D\|} + \Gamma_i + \Gamma_f$; $\alpha = e^2/\hbar c$ is the fine-structure constant. The expressions (4) and (5) are written under the assumption that $\Delta\varepsilon_{HSF} \leq \Delta\omega_2$, and the interference of the amplitudes of the corresponding transitions has been taken into account in them. For other aspects one can see [8-61].

The formulas (4) and (5) determine the difference of the probabilities of population of $2s_{1/2}$ states with different projections M of the nuclear spin:

$$\Delta w = w\left(M = \frac{1}{2}\right) - w\left(M = -\frac{1}{2}\right) = -\frac{8}{27}(2\pi)^2 \alpha |\langle 2s|r|3p\rangle|^2 \Delta\varepsilon_{HSF}$$

$$\times \left[\frac{|V^{(1)}_{p/n_1}|^2}{(\varepsilon - \varepsilon_{01})^2 + \tilde{\Gamma}^2_{n_1}/4} \frac{dI}{d\omega_2}\bigg|_{\omega_2} - \frac{1}{2}\frac{|V^{(1)}_{p/n_2}|^2}{(\varepsilon - \varepsilon_{02})^2 + \tilde{\Gamma}^2_{n_2}/4} \frac{dI}{d\omega_2}\bigg|_{\omega_2}\right]. \qquad (6)$$

As we should expect, when the condition $\Delta\varepsilon_{HSF} \leq \Delta\omega_2$ is fulfilled, the probability difference (6) has turned out to be proportional to the energy $\Delta\omega_{HSF}$ of the hyperfine splitting of the F = 1 and F = 0 components of the total angular momentum of the $2s_{1/2}$ state of the atom and is determined by the values of the derivatives of the spectral density of the radiation in the second wave at the frequencies of the transitions $3p_{3/2} \to 2s_{1/2}$ (F = 1) and $3p_{1/2} \to 2s_{1/2}$ (F =1).